\documentclass[aps,preprint,showpacs,amsmath,amssymb]{revtex4-1}
\usepackage{graphicx}
\usepackage{dcolumn}% Align table columns on decimal point
\usepackage{bm}% bold math
\usepackage{hyperref}
\usepackage{subfigure}

\begin{document}

\title{Freezing dynamics of genuine entanglement and loss of genuine nonlocality under collective dephasing}
\author{Mazhar Ali}
\affiliation{Department of Electrical Engineering, Faculty of Engineering, Islamic University Madinah, 107 Madinah, Saudi Arabia}

\begin{abstract}
We study the dynamics of genuine multipartite entanglement for quantum systems upto four qubits interacting with 
general collective dephasing process. Using a computable entanglement monotone for multipartite systems, we observe the 
feature of freezing dynamics of genuine entanglement for three and four qubits entangled states. We compare the dynamics with that of random states 
and find that most states exibit this feature. We then study the effects of collective dephasing on genuine nonlocality and find out that although 
quantum states remain genuinely entangled yet their genuine nonlocality is lost in a finite time. We show the sensitivity of asymptotic states being 
genuinely entangled by mixing white noise.  
\end{abstract}


\maketitle

%\keywords{Multipartite genuine entanglement; Decoherence; Genuine Nonlocality}

%\date{\today}h

%\pacs{03.65.Yz, 03.65.Ud, 03.67.Mn}

%%%%%%%%%%%%%%%%%%%%%%%%%%%%%%%%%%%%%%%
\section{Introduction}\label{S-intro}
%%%%%%%%%%%%%%%%%%%%%%%%%%%%%%%%%%%%%%%

Quantum entanglement and nonlocality are features of quantum world which has attracted lot of interest to develop a theory of its own 
\cite{Horodecki-RMP-2009,gtreview,Brunner-RMP}. Due to growing efforts for an experimental realization of devices utilizing these features, 
it is essential to study the effects of noisy environments on entanglement and nonlocality. 
Such studies are an active area of research \cite{Aolita-review} and several authors have studied decoherence effects on quantum correlations for 
both bipartite and multipartite systems
\cite{Yu-work,lifetime,Aolita-PRL100-2008,bipartitedec,Band-PRA72-2005,lowerbounds,Lastra-PRA75-2007,Guehne-PRA78-2008,Lopez-PRL101-2008,Ali-work,Weinstein-PRA85-2012,Ali-JPB-2014}. 

One specific type of noise dominant in experiments on trapped atoms is caused by intensity fluctuations of electromagnetic fields which leads 
to collective dephasing process. The detrimental effects of collective dephasing noise on entanglement have been studied 
\cite{Yu-CD-2002,AJ-JMO-2007,Li-EPJD-2007,Karpat-PLA375-2011,Ali-PRA81-2010,Liu-arXiv,Ali-EJPD-2017}, however all these previous studies were 
restricted to a special orientation (z-axes) of the field. Recently, a more general approach has been worked out 
\cite{Carnio-PRL-2015,Carnio-NJP-2016}, where the authors addressed an arbitrary orientation of field. This general approach revealed an 
interesting feature of its dynamical nature. It was shown by taking a specific two qubits state \cite{Carnio-PRL-2015} that certain orientation 
of field give rise to stationary asymptotic entanglement. Also for multipartite $W$ state, it was found using certain inequalities that 
asymptotic state is entangled. In our work, we extend the previous studies to include some other classes of multipartite entangled states. 
Recent progress in the theory of multipartite entanglement has enabled us to study decoherence effects on actual multipartite 
entanglement and not on entanglement among bipartitions. In particular, the ability to compute genuine negativity for multipartite systems 
has eased this task \cite{Bastian-PRL106-2011}.
In addition, in order to better judge the dynamical behaviour of a state, we need to compare its dynamics with dynamics of random states. 
Our analysis suggests that freezing entanglement phenomenon is a generic feature of the dynamical process, as almost all states exhibit this phenomenon. 

Another concept related to non-classical correlations is quantum nonlocality. This feature says that the predictions made using quantum 
mechanics cannot be simulated by a local hidden variable model. The presence of nonlocal correlations can be shown via violation of Bell 
inequalities \cite{Bell-Phys-1964}. The pure entangled states violate a Bell inequality, whereas mixed entangled states may not do 
so \cite{Gisin-Werner-1991}. However, all entangled states do exhibit some kind of hidden nonlocality \cite{Liang-PRA86-2011}. The well known 
Clauser-Horne-Shimony-Holt (CHSH) inequality \cite{CHSH-1969} for two qubits has been studied in the presence of decoherence both in 
theory \cite{Mazzola-PRA81-2010}, and experimentally \cite{Xu-PRL-2010}. The extension of CHSH inequality for multipartite quantum systems 
has received considerable attention \cite{Mermin-PRL-1990,Ardehali-PRA-1992,Collins-PRL-2002,Bancal-PRL-2011}, however Svetlichny discovered 
the first method to detect genuine multipartite nonlocality \cite{Svetlichny-PRD-1987}. Violations of some of these inequalities in experiments 
have also been reported \cite{Pan-expMBI,Bastian-PRL104-2010}. 
Several investigations of nonlocality of multipartite quantum states under decoherence have been carried out \cite{NLD}.
We also study the decoherence effects on genuine nonlocality quantified by 
Svetlichny inequality. We find that although the quantum states might remain genuinely entangled, nevertheless they lose their genuine nonlocality 
in finite time. This observation is similar to two qubits case  where there are instances when the states remain entangled whereas nonlocality is 
lost in finite time.

This paper is organized as follows. In section \ref{Sec:Model}, we briefly describe our physical model and equations of motion. We review concepts of 
genuine entanglement and genuine nonlocality in section \ref{Sec:GME}. We present main results in section \ref{Sec:results}. 
Finally we conclude the work in section \ref{Sec:conc}. 

%%%%%%%%%%%%%%%%%%%%%%%%%%%%%%%%%%%%%%%%%%%%%%%%%%%%%%%%%%%%%%%%%%%%%%%%%%%%%%%%%%%%%%%%%%%%%%%%%%%%%%%%%%%%%%
\section{Open-system dynamics of multi-qubits under collective dephasing} \label{Sec:Model}
%%%%%%%%%%%%%%%%%%%%%%%%%%%%%%%%%%%%%%%%%%%%%%%%%%%%%%%%%%%%%%%%%%%%%%%%%%%%%%%%%%%%%%%%%%%%%%%%%%%%%%%%%%%%%%

In this section, we describe the physical model and equations of motion governing our system of interest. We consider our qubits as 
atomic two-level systems with energy splitting $\hbar \omega$. The splitting is controlled by a homogeneous magetic field. 
The Hamiltonian for a single atom is given by 
\begin{eqnarray}
\hat{H}_\omega = \frac{\hbar \, \omega}{2} \, \vec{n} \cdot \vec{\sigma} \,,
\label{Eq:Hamil}
\end{eqnarray}
where $\vec{n} = n_x \hat{x} + n_y \hat{y} + n_z \hat{z}$ is the orientation of magnetic field and 
$\vec{\sigma} = \sigma_x \hat{x} + \sigma_y \hat{y} + \sigma_z \hat{z}$ is the vector of standard Pauli matrices. 
This time independent Hamiltonian generates the propagator 
\begin{eqnarray}
U_\omega(t) = {\rm e}^{- i \, H_\omega \, t/\hbar } = {\rm e}^{- i \, \omega \, t/2 \, \, \, {\bf n} \cdot {\bf \sigma} } \,. 
\label{Eq:Uw}  
\end{eqnarray}
We can introduce a pair of orthogonal projectors 
\begin{eqnarray}
\Lambda_\pm = \frac{I_2 \pm {\bf n}\cdot {\bf \sigma}}{2} \, ,
\label{Eq:TE}
\end{eqnarray}
to write the propagtor in terms of them. Let us consider $N$ non-interacting atoms (qubits), so that the propagator for these collection of atoms 
can be written as \cite{Carnio-PRL-2015} 
\begin{eqnarray}
U_\omega(t)^{\otimes \, N} =& \, (e^{-i \, \omega \, t} \, \Lambda_+ \, + e^{i \, \omega \, t} \, \Lambda_-)^{\otimes \, N} \nonumber \\&
= \sum_{j = 0}^N \, e^{i\, \omega \,t (j-N/2) } \, \Theta_j \, ,
\label{Eq:UwN}
\end{eqnarray}
where the operators $\Theta_j$ are defined as 
\begin{eqnarray}
\Theta_j = \frac{1}{j! \, (N-j)!} \, \sum_{s \in \sum_N} \, V_s \, \big[ \Lambda_-^{\otimes \, j} 
\otimes \Lambda_+^{\otimes \, N-j} \big] \, V_s^\dagger \, , 
\label{Eq:Theta}
\end{eqnarray}
where $\sum_N$ represents the symmetric group and $V_s$ are the permutations in operator space of $N$ qubits. 

As there are fluctuations in the magetic field strength, the integration over it will induce a probability distribution $p(\omega)$ of 
characteristic energy splitting. Therefore the time evolution of the combined state of $N$ atoms can be written as \cite{Carnio-PRL-2015}
\begin{eqnarray}
\rho(t) = \int p(\omega) \, U_\omega(t)^{\otimes \, N} \, \rho(0) \, U^\dagger_\omega(t)^{\otimes \, N} \, d\omega \,.
\end{eqnarray}
In writing this equation, we have assumed that the field fluctuations occur on time scales which are longer than the times over which the combined 
state of $N$ atoms evolve under unitary propagator $U_\omega(t)^{\otimes N}$. Substituting the above derived format for the unitary propagator, we can 
write the time evolved state as 
\begin{eqnarray}
\rho(t) = \sum_{j,k = 0}^N \, M_{jk}(t) \, \Theta_j \, \rho(0) \, \Theta_k \, ,
\label{Eq:TES}
\end{eqnarray}
where $M_{jk}(t)$ are elements of the Toeplitz matrix $M(t)$, which can be obtained by the relation $M_{jk}(t) = \phi[(j-k)t]$, where $\phi(t)$ is 
the characteristic function of the probability distribution $p(\omega)$, defined as
\begin{eqnarray}
 \phi(t) = \int \, p(\omega) \, e^{i \, \omega \, t} \, d\omega \, .
\end{eqnarray}
It has been demonstrated that time evolution form Eq.(\ref{Eq:TES}) is both trace preserving and positivity preserving \cite{Carnio-PRL-2015}. 
This means that the dynamical process is a valid equation of motion. In order to study the exact behaviour of multipartite quantum states, 
it is convenient to obtain an exact expression for state $\rho(t)$, in terms of a spectral distribution $p(\omega)$ characterizing the fluctuations. 
As an example, we take the Lorentzian distribution also known as Cauchy distribution, defined as
\begin{equation}
 p(x) = \frac{\gamma^2}{\pi \, \gamma \, \big[ (x - x_0)^2 + \gamma^2 \big]}\,.
\end{equation}
For standard Cauchy distribution, the characteristic function turns out to be 
\begin{equation}
 \phi(t) = \int \, p(x) \, e^{i \, x \, t} \, dx = e^{-|t|} \,.
\end{equation}

In this work, we restrict ourselves with upto four qubits. Let us first consider the simplest case of two qubits. The Toeplitz matrix for two qubits  
with Lorentzian distribution can be obtained straightforwardly as
\begin{eqnarray}
M(t) = \left( 
\begin{array}{ccc}
1 & e^{-t} & e^{-2 \, t} \\ 
e^{-t} & 1 & e^{-t} \\ 
e^{-2 \, t} & e^{-t} & 1 
\end{array}
\right)\,.
\label{Eq:Mt2Q}
\end{eqnarray}
Here $t$ denotes the usual dimensionless quantity $\Gamma t$. The operators $\Theta_j$ for two qubits are given as \cite{Carnio-NJP-2016} 
\begin{eqnarray}
\Theta_0 =& \, \Lambda_+ \otimes \Lambda_+ \nonumber \\  
\Theta_1 =& \, \Lambda_+ \otimes \Lambda_- + \Lambda_- \otimes \Lambda_+ \nonumber \\
\Theta_2 =& \, \Lambda_- \otimes \Lambda_- \,.
\label{Eq:Thta3}
\end{eqnarray}
The time evolution of an arbitrary initial state can be obtained straightforwardly. One important and interesting feaure of this dynamics is the 
observation that in general there are no decoherence free spaces (DFS) in this noisy model except for $\vec{n} = (0,0,1)^T$ which is the case mostly 
studied \cite{Yu-CD-2002,AJ-JMO-2007,Li-EPJD-2007,Karpat-PLA375-2011,Ali-PRA81-2010,Liu-arXiv} for bipartite and multipartite systems. 
For other two special directions $\vec{n} = (1,0,0)^T$ and $\vec{n} = (0,1,0)^T$, it may 
happen that some quantum states are completely invariant. Obviously all entangled states residing in DFS and/or invariant under certain 
directions of magnetic field maintain their correlation properties. However, there exist another non-trivial and interesting dynamics which is so called 
{\it time-invariant entanglement} such that the quantum states are changing at every instance however their entanglement remains constant. Such observation 
was initially made for qubit-qutrit systems \cite{Karpat-PLA375-2011} and later on for a specific family which is so called Bell-diagonal states of two 
qubits \cite{Liu-arXiv}. Recently, we have studied this peculier {\it time-invariant} entanglement feature for multipartite systems and have shown that
genuine entanglement also exhibits this {\it time-invariant} behaviour \cite{Ali-EJPD-2017}. However, we noticed that {\it time-invariant} entanglement 
occurs only for special orientation of field, that is, for $\vec{n} = (0,0,1)^T$ as an example, where we have decoherence free subspaces. 

The more general description of collective dephasing for an arbitrary direction of vector $\vec{n}$ was worked out only 
recently \cite{Carnio-PRL-2015}. For $\vec{n}$ other than special orientations mentioned in previous paragraph, there are no decoherence free subspaces but 
there is another interesting feature of entanglement dynamics which is completely different than {\it time-invariant} feature, namely the dynamical 
{\it freezing} of entanglement. It happens that initial quantum states lose their entanglement upto some specific value, and afterwards exhibit 
frozen dynamics of entanglement whereas the quantum states are changing at every instance. A concrete example was presented for a specific quantum state 
of two qubits \cite{Carnio-PRL-2015} and predictions were made for multipartite quantum states. 

We now move to three qubits where the Toeplitz matrix is given as
\begin{eqnarray}
M(t) = \left( 
\begin{array}{cccc}
1 & e^{-t} & e^{-2 \, t} & e^{-3 \, t} \\ 
e^{-t} & 1 & e^{-t} & e^{-2 \, t} \\ 
e^{-2 \, t} & e^{-t} & 1 & e^{-t} \\
e^{-3 \, t} & e^{-2 \, t} & e^{-t} & 1
\end{array}
\right)\,, 
\label{Eq:Mt}
\end{eqnarray}
and the operators $\Theta_j$ for three qubits are given as 
\begin{eqnarray}
\Theta_0 =& \, \Lambda_+ \otimes \Lambda_+ \otimes \Lambda_+ \, \nonumber \\  
\Theta_1 =& \, \Lambda_+ \otimes \Lambda_+ \otimes \Lambda_- + \Lambda_+ \otimes \Lambda_- \otimes \Lambda_+ 
+ \Lambda_- \otimes \Lambda_+ \otimes \Lambda_+ \nonumber \\
\Theta_2 =& \, \Lambda_+ \otimes \Lambda_- \otimes \Lambda_- + \Lambda_- \otimes \Lambda_+ \otimes \Lambda_-  + 
\Lambda_- \otimes \Lambda_- \otimes \Lambda_+ \nonumber \\
\Theta_3 =& \, \Lambda_- \otimes \Lambda_- \otimes \Lambda_- \,.
\label{Eq:Thta3}
\end{eqnarray}
It is straightforward to obtain an analytical expression for the time evolution of an arbitrary initial state of three qubits. However, due to 
presence of parameters $n_i$, the resulting density matrix is quite cumbersome to present here. One important observation is the fact that the 
initial states do not preserve their initial form, for an example, the $X$ states do not remain in $X$ form and many other matrix elements become
non-zero even they were zero initially. 

The Toeplitz matrix and operators for four qubits are simple extension of the above mentioned cases so we do not write their exact form here. 

%%%%%%%%%%%%%%%%%%%%%%%%%%%%%%%%%%%%%%%%%%%%%%%%%%%%%%%%%%%%%%%%%%%%%%%%%
\section{Genuine multipartite entanglement and genuine nonlocality} 
\label{Sec:GME}
%%%%%%%%%%%%%%%%%%%%%%%%%%%%%%%%%%%%%%%%%%%%%%%%%%%%%%%%%%%%%%%%%%%%%%%%%

In this section, we briefly review the basic definitions for genuine multipartite entanglement and genuine nonlocality. 
We discuss the main ideas by considering three parties $A$, $B$, and $C$, generalization to more parties being straightforward. 
A state is called separable with respect to some bipartition, say, $A|BC$, if it is a mixture of product states with respect 
to this partition, that is, 
$\rho = \sum_j \, p_j \, |\psi_A^j \rangle\langle \psi_A^j| \otimes |\psi_{BC}^j \rangle\langle \psi_{BC}^j|$, 
where the $p_j$ form a probability distribution. We denote these states as $\rho_{A|BC}^{sep}$. 
Similarly, we can define separable states for the two other bipartitions, $\rho_{B|CA}^{sep}$ and $\rho_{C|AB}^{sep}$. 
Then a state is called biseparable if it can be written as a mixture of states which are separable with respect 
to different bipartitions, that is 
\begin{eqnarray}
 \rho^{bs} = \tilde{p}_1 \, \rho_{A|BC}^{sep} + \tilde{p}_2 \, \rho_{B|AC}^{sep} + \tilde{p}_3 \, \rho_{C|AB}^{sep}\,,
\end{eqnarray}
with $\tilde{p}_1 +\tilde{p}_2 +\tilde{p}_3 = 1$.
Finally, a state is called genuinely multipartite entangled if it is not biseparable. In the rest of this paper, we always mean genuine
multipartite entanglement when we talk about entanglement. 

Genuine entanglement can be detected and characterized \cite{Bastian-PRL106-2011} by a technique based on positive partial transpose 
mixtures (PPT mixtures). A two-party state $\rho = \sum_{ijkl} \, \rho_{ij,kl} \, |i\rangle\langle j| \otimes |k\rangle\langle l|$ is PPT if 
its partially transposed matrix $\rho^{T_A} = \sum_{ijkl} \, \rho_{ji,kl} \, |i\rangle\langle j| \otimes |k\rangle\langle l|$ is positive semidefinite. 
The separable states are always PPT \cite{peresppt} and the set of separable states with respect to some partition 
is therefore contained in a larger set of states which has a positive partial transpose for that bipartition. 

Denoting PPT states with respect to fixed bipartition by $\rho_{A|BC}^{PPT}$, $\rho_{B|CA}^{PPT}$, 
and $\rho_{C|AB}^{PPT}$, we call a state as PPT-mixture if it can be written as
\begin{eqnarray}
\rho^{PPTmix} = q_1 \, \rho_{A|BC}^{PPT} + q_2 \, \rho_{B|AC}^{PPT} + q_3 \, \rho_{C|AB}^{PPT}\,.
\end{eqnarray}
As any biseparable state is a PPT mixture, therefore any state which is not 
a PPT mixture is guaranteed to be genuinely multipartite entangled. The main advantage of considering PPT mixtures instead 
of biseparable states comes from the fact that PPT mixtures can be fully characterized by the method of semidefinite programming 
(SDP), a standard  method in convex optimization \cite{sdp}. Generally the set of PPT mixtures is a very good approximation to the 
set of biseparable states and delivers the best known separability criteria for many cases; however, there are multipartite entangled 
states which are PPT mixtures \cite{Bastian-PRL106-2011}. The description of SDP and genuine negativity is described in details in 
Ref.\cite{Bastian-PRL106-2011}.   
In order to quantify genuine entanglement, it was shown \cite{Bastian-PRL106-2011} that for the following optimization problem 
\begin{eqnarray}
\min {\rm Tr} (\mathcal{W} \rho)
\end{eqnarray}
with constraints that for all bipartition $M|\bar{M}$
\begin{eqnarray}
\mathcal{W} = P_M + Q_M^{T_M},
 \quad \mbox{ with }
 0 \leq P_M\,\leq 1 \mbox{ and }
 0 \leq  Q_M  \leq 1\, 
\end{eqnarray}
the negative witness expectation value is multipartite entanglement monotone. The constraints just state that the considered operator $\mathcal{W}$ is a 
decomposable entanglement witness for any bipartition. Since this is a semidefinite program, the minimum can be efficiently computed and the optimality 
of the solution can be certified \cite{sdp}. We denote this measure by $E(\rho)$ or $E$-monotone in this paper. For bipartite systems, this monotone is 
equivalent to {\it negativity} \cite{Vidal-PRA65-2002}. For a system of qubits, this measure is bounded by  $E(\rho) \leq 1/2$ \cite{bastiangraph}.

For a brief description of genuine nonlocality, consider that each party can perform a measurement $X_j$ with result $a_j$ for $j = A,B,C$. The 
joint probability distribution $P(a_A a_B a_C|X_A X_B X_C)$ may exhibit different notions of nonlocality. It may be that it cannot be written 
in local form as
\begin{equation}
 P(a_A a_B a_C|X_A X_B X_C) = \int p_\lambda d\lambda P_A(a_A|X_A \lambda) \, P_B(a_B|X_B \lambda)\, P_C(a_C|X_C \lambda) \,,
\label{Eq:PLV}
 \end{equation}
where $\lambda$ is a shared local variable. Such nonlocality can be tested by standard Bell inequalities and it can not capture the 
genuine nonlocality. As an example consider that parties $A$ and $B$ are nonlocally correlated but uncorrelated from party $C$. It is still 
possible that $P$ cannot be written as Eq.(\ref{Eq:PLV}), although the system has no genuine tripartite nonlocality \cite{Bancal-PRL-2011}. 
Genuine nonlocality can be detected if one makes sure that $P$ cannot be written as 
\begin{equation}
P_G(a_A a_B a_C|X_A X_B X_C) = \sum_{m = 1}^{3} \, p_m \, \int d\lambda \rho_{ij}(\lambda) P_{ij}(a_i a_j|X_i X_j \lambda) \, P_m(a_m|X_m \lambda)\, \,,
\label{Eq:GLV}
 \end{equation}
that is, $P$ cannot be reproduced by local means even if any two of parties come together and act jointly to produce bipartite nonlocal 
correlations with probability distribution $\rho_{ij}(\lambda)$, where $ij$ denotes for all possible partitions. By focusing on the possible that 
each party $j$ is allowed to two measurements $X_j$ and $X'_j$ with outcomes $a_j$ and $a'_j$ such that $a_j, \, a'_j \in \{-1, 1\}$, one of 
possible form of Svetlichny \cite{Svetlichny-PRD-1987} inequality is given as  
\begin{eqnarray}
S_3 =  \big( a_A a_B a_C + a_A a_B a'_C + a_A a'_B a_C + a'_A a_B a_C - a'_A a'_B a'_C \nonumber \\  
 -  a'_A a'_B a_C - a'_A a_B a'_C - a_A a'_B a'_C \big) \leq 4 \,. 
\end{eqnarray}
By associating optimal measurements $M$ (details in next section) with each term \cite{AJ-FP-2009}, we can write the expectation value of this operator as  
\begin{eqnarray}
 \langle S \rangle =  {\rm Tr} \big[ \big( M_A M_B M_C + M_A M_B M'_C + M_A M'_B M_C + M'_A M_B M_C \nonumber \\ - M'_A M'_B M'_C 
 -  M'_A M'_B M_C - M'_A M_B M'_C - M_A M'_B M'_C \big) \rho(t) \big]\,. 
\end{eqnarray}
It is well known that the $n$-partite quantum state $\rho(t)$ exibits genuine multipartite nonlocality if $|\langle S \rangle| > 2^{n-1}$. Hence for three
qubits, the violation of Svetlichny inequality guarantees the presence of genuine tripartite nonlocality. It is interesting to note the similarity of ideas
between detecting genuine entanglement and genuine nonlocality.

%%%%%%%%%%%%%%%%%%%%%%%%%%%%%%%%%%%%%%%
\section{Results} \label{Sec:results}
%%%%%%%%%%%%%%%%%%%%%%%%%%%%%%%%%%%%%%%

In this section, we present our main results for various initial states of three and four qubits. Our choice of initial states are defined as
\begin{equation}
\rho = \alpha |\Psi\rangle\langle \Psi| + \frac{1-\alpha}{2^N} \, I_{N^2} \, ,
\label{Eq:IS}
\end{equation}
where $0 \leq \alpha \leq 1$, $|\Psi\rangle$ can be any genuine entangled random or specific pure state of three and/or four qubits and $I_{N^2}$ is 
the identity matrix for $N$ qubits with dimension $N^2$. 
We are interested in several families of states in this work. Two important families of states, namely the GHZ states and the W states 
for $N$ qubits are given as  
\begin{eqnarray}
|GHZ_N \rangle &=& \frac{1}{\sqrt{2}}(|00...0\rangle + |11...1\rangle), \nonumber \\
|W_N\rangle &=& \frac{1}{\sqrt{N}}(|00...001\rangle + |00...010\rangle + \ldots + |10...000\rangle).
\label{Eq:GHZ3Qb1}
\end{eqnarray}
GHZ state has always maximum value of monotone, that is, $E(|GHZ_N\rangle\langle GHZ_N|) = 1/2$, whereas for the W state, numerical 
value depends on the number of qubits. For three qubits $E(|W_3\rangle\langle W_3|) \approx 0.443$ and for four qubits 
$E(|W_4\rangle\langle W_4|) \approx 0.366$.

Several interesting states for four qubits are Dicke state $|D_{2,4}\rangle$, the singlet state $|\Psi_{S,4}\rangle$, the cluster 
state $|CL\rangle$ and the so-called $\chi$-state $|\chi_4\rangle$, given as 
\begin{eqnarray}
|D_{2,4} \rangle &=& \frac{1}{\sqrt{6}} [ |0011 \rangle + |1100\rangle + |0101 \rangle + |0110\rangle + |1001 \rangle 
+ | 1010\rangle] \,,  \nonumber \\
|\Psi_{S,4}\rangle &=& \frac{1}{\sqrt{3}} [ |0011 \rangle + |1100\rangle - \frac{1}{2} ( \, |0101 \rangle + |0110\rangle  
+ |1001 \rangle + |1010\rangle)] \,, \nonumber \\
|CL \rangle &=& \frac{1}{2} [|0000 \rangle + |0011\rangle + |1100 \rangle - |1111\rangle], \nonumber \\
|\chi_4 \rangle &=& \frac{1}{\sqrt{6}} [\sqrt{2} | 1111 \rangle + |0001 \rangle + |0010\rangle + |0100 \rangle + |1000\rangle], 
\end{eqnarray}
respectively. All these states are maximally entangled with respect to multipartite negativity, 
$E (|D_{2,4}\rangle \langle D_{2,4}|) = E (|\Psi_{S,4}\rangle \langle \Psi_{S,4}|) = E (|CL\rangle \langle CL|) 
= E (|\chi_{4}\rangle\langle \chi_4|) = 1/2$. These states along with their properties are discussed in Ref.~\cite{gtreview}.

We also study the behaviour of random states which are generated as follows \cite{Toth-Arxiv}: First, we generate a vector such that 
both the real and the imaginary parts of the vector elements are Gaussian distributed random numbers with a zero mean and unit 
variance. Second we normalize the vector. It is easy to prove that the random vectors obtained this way are equally distributed 
on the unit sphere \cite{Toth-Arxiv}. We stress that we generate random pure states in the 
global Hilbert space of three qubits, so the unit sphere is not the Bloch ball.

\subsection{Three qubits}

\begin{figure}[t!]
\centering
\scalebox{2.20}{\includegraphics[width=5.5cm]{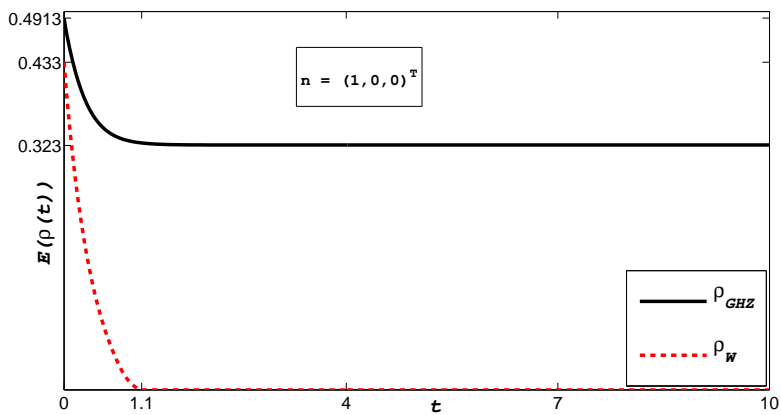}}
\caption{$E$-monotone is plotted against dimensionless parameter $t$ for three qubits GHZ and W states in Eq.~(\ref{Eq:IS}). 
The black solid line is for GHZ mixture and red dashed line is for mixture of W state. We have taken $\alpha = 0.99$ and $\vec{n} = (1,0,0)^T$ to plot 
these curves. See text for more description.}
\label{FIG:WGHZn1}
\end{figure}

Let us first take mixtures of GHZ and W states in Eq.~(\ref{Eq:IS}) with $\alpha = 0.99$. 
Figure (\ref{FIG:WGHZn1}) depicts the genuine negativity for these states, where we have taken $\vec{n} = (1,0,0)^T$ to 
obtain these curves. It can be seen that initially entanglement starts decaying for $\rho_{GHZ}(t)$ state, however after a while, it 
reaches a value of $\approx 0.323$ and after that the states exhibits freezing dynamics for genuine entanglement. On the other hand, genuine 
negativity for similar mixture of W state becomes zero after a short time and does not exhibit such behaviour for this specific direction of magnetic 
field. 

\begin{figure}[t!]
\centering
\scalebox{2.20}{\includegraphics[width=5.5cm]{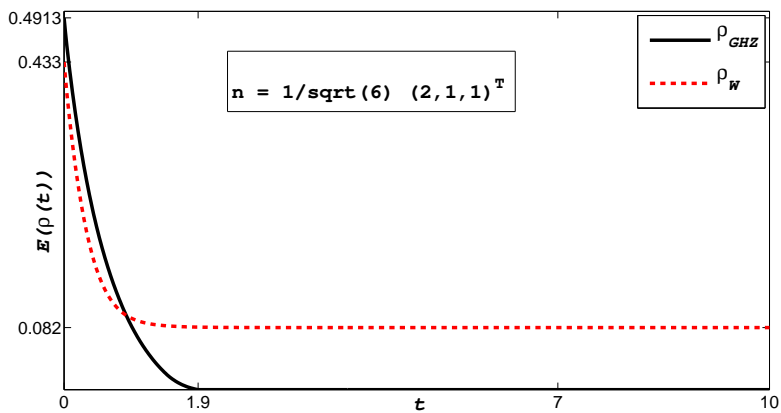}}
\caption{Genuine negativity is plotted against dimensionless parameter $t$ for three qubits GHZ and W states in Eq.~(\ref{Eq:IS}). 
The black solid line is for GHZ mixture and red dashed line is for mixture of W state. We have taken $\alpha = 0.99$ and 
$\vec{n} = (2,1,1)^T/\sqrt{6}$ to plot these curves. See text for more description.}
\label{FIG:WGHZn2}
\end{figure}
For another choice of $\vec{n} = (2,1,1)^T/\sqrt{6}$, we observe quite opposite behaviour of genuine negativity for GHZ and W state. 
In Figure~(\ref{FIG:WGHZn2}) we plot genuine negativity for same mixtures of GHZ and W states by taking $\alpha = 0.99$. 
We can see that now GHZ state (black solid curve) looses its genuine negativity at a short time $ t \approx 1.9$ whereas W state (red dashed line) 
first looses its entanglement to value $\approx 0.082$ and after that exhibits freezing dynamics. 

\begin{figure}
\scalebox{1.75}{\includegraphics[width=1.55in]{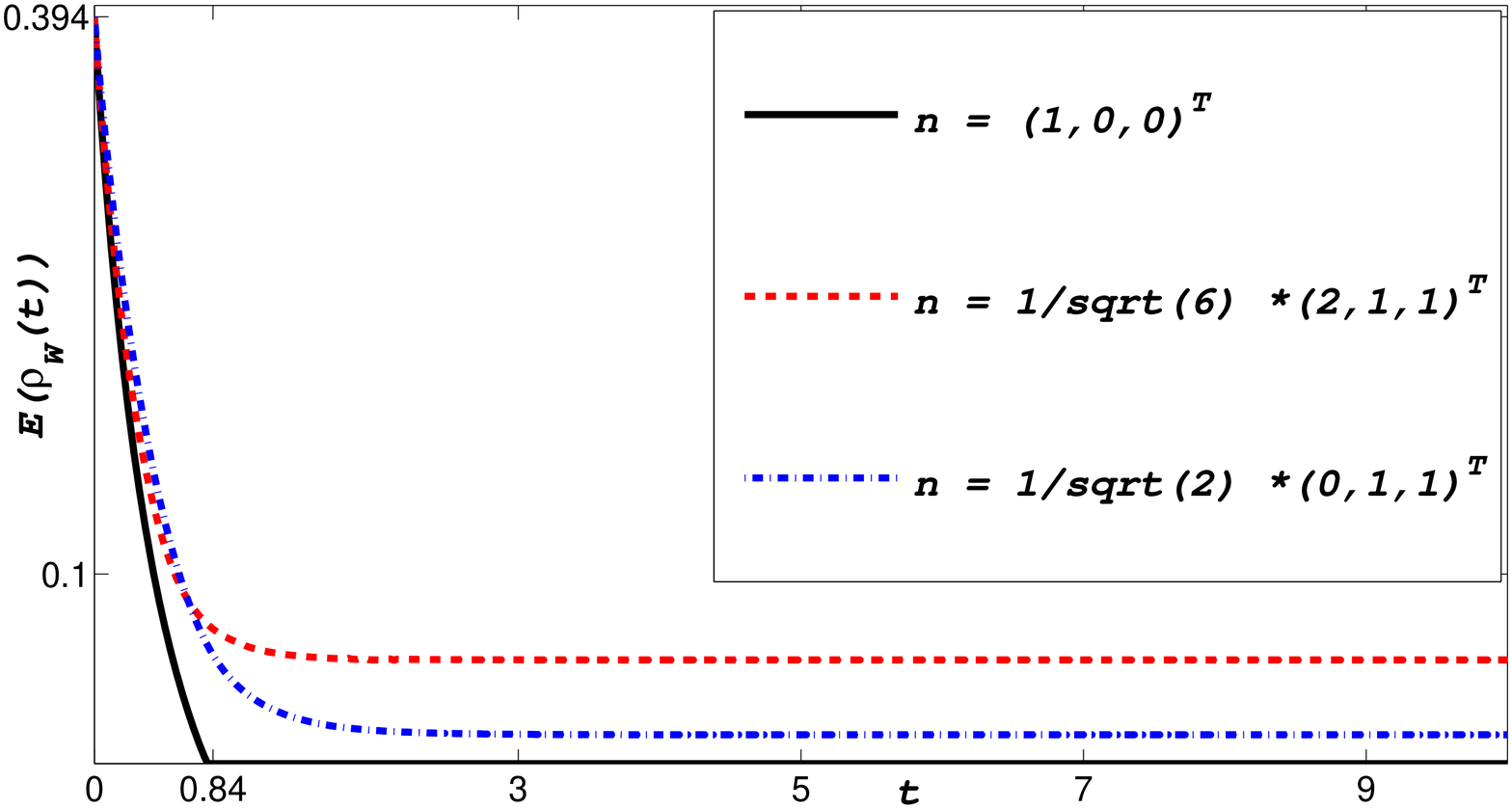}}
\scalebox{1.75}{\includegraphics[width=1.55in]{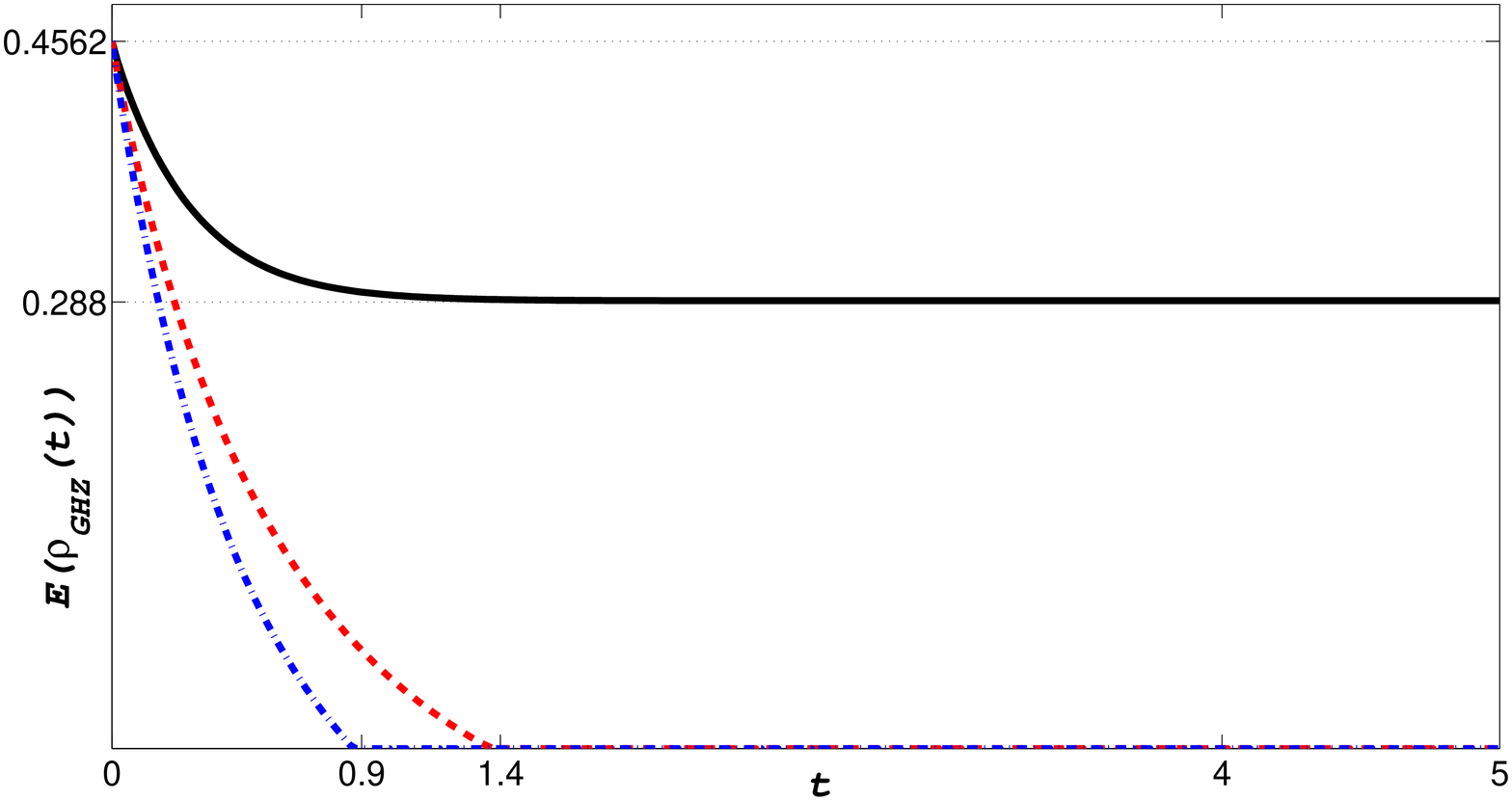}}
\centering
%\subfigure[]{\includegraphics[width=8.1cm]{FigWEFLD1}}
%\subfigure[]{\includegraphics[width=8.1cm]{FigGHZEFLD1}}
\caption{(a)Genuine negativity is plotted against dimensionless parameter $t$ for mixture of three qubits W state with $\alpha =0.95$ for various 
settings of $\vec{n}$. (b) Same caption and color codes as for part (a) but for mixture of GHZ state. See text for details.}
\label{FIG:WEFn123}
\end{figure}

In order to check how freezing dynamics of entanglement is effected by white noise, we now increase the percentage of noise from $1\%$ to 
$5\%$ by setting $\alpha = 0.95$. In Figure~(\ref{FIG:WEFn123}), we plot genuine negativity for 
mixtures of W and GHZ states for various settings of $\vec{n}$. In Figure~(\ref{FIG:WEFn123}a) we see that for W states, the increament of white noise 
has brought the loss of entanglement earlier as expected for $\vec{n} = (1,0,0)^T$, whereas for other two settings of $\vec{n}$, we have 
freezing dynamics. Figure~(\ref{FIG:WEFn123}b) results the genuine negativity for GHZ state for three same settings of $\vec{n}$. We find that except for 
$\vec{n} = (1,0,0)^T$, entanglement is lost at finite time and freezing dynamics is again exhibited however at a slight lower value than earlier due to more 
noise.

% In Figure (\ref{FIG:GHZEFn123}), we plot genuine negativity for various choices of $\vec{n}$ and $\alpha = 0.95$. The black solid line is for 
% $\vec{n} = (1,0,0)^T$ and exhibits freezing dynamics of genuine entanglement at $E(\rho) \approx 0.288$, slightly lower as compared with 
% $\alpha = 0.99$. The red dashed line is for $\vec{n} = (2,1,1)^T/\sqrt{6}$ and comes to an end at $\approx 1.4$. The curve for 
% $\vec{n} = (0,1,1)^T/\sqrt{2}$ (blue dashed dotted line) also becomes zero at $\approx 0.9$.   
% \begin{figure}[t!]
% \centering
% \scalebox{2.20}{\includegraphics[width=5.5cm]{FigGHZEFLD1}}
% \caption{Genuine negativity is plotted against dimensionless parameter $t$ for three qubits $\rho_{GHZ}$ states Eq.(\ref{}) for 
% $\vec{n} = (1,0,0)^T$ (black solid line),  $\vec{n} = (2,1,1)^T/\sqrt{6}$ (red dashed line), and $\vec{n} = (0,1,1)^T/\sqrt{2}$ (blue dashed dotted 
% line). We take $\alpha = 0.95$.}
% \label{FIG:GHZEFn123}
% \end{figure}
\begin{figure}
\scalebox{1.75}{\includegraphics[width=1.55in]{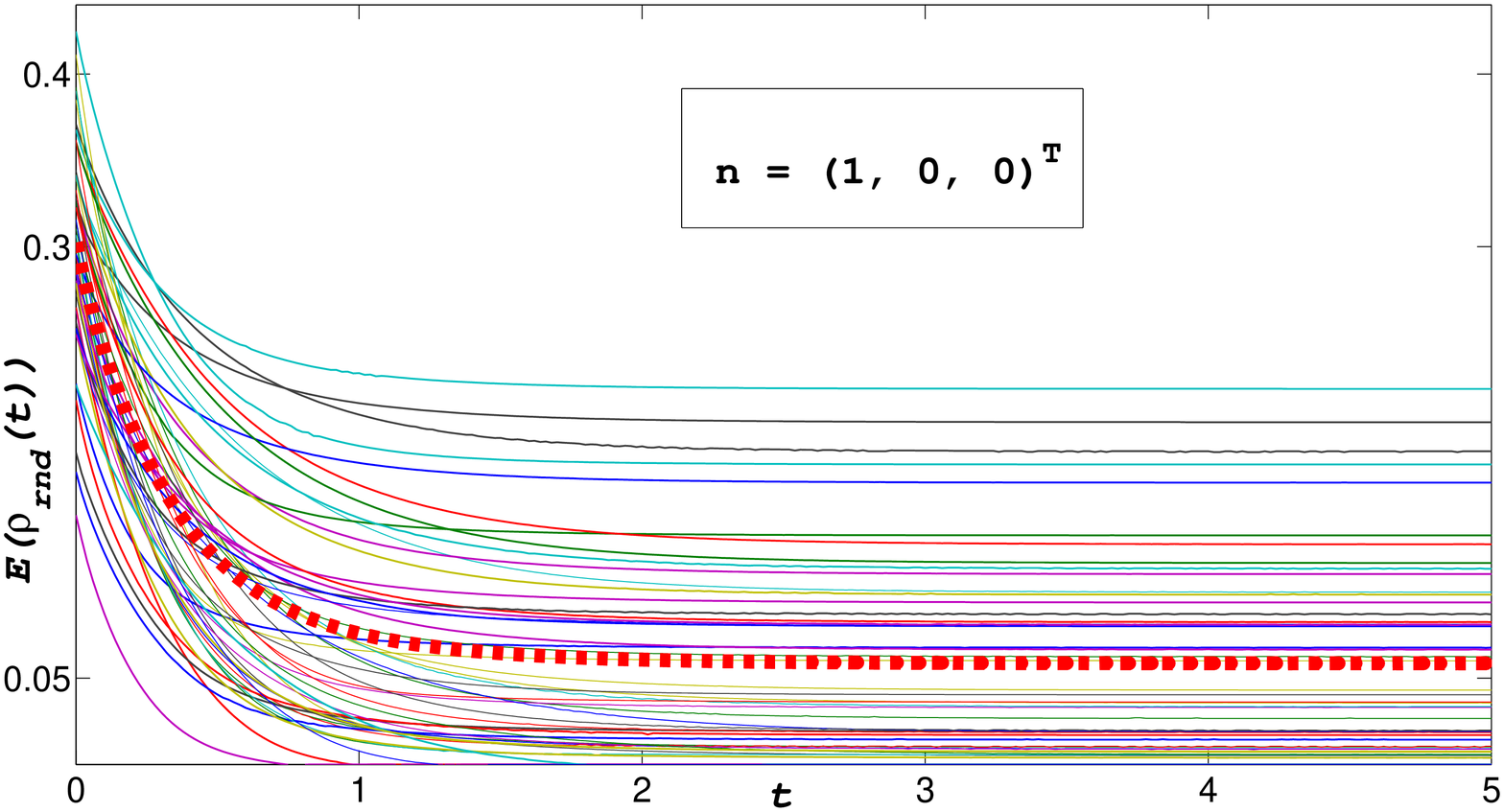}}
\scalebox{1.75}{\includegraphics[width=1.55in]{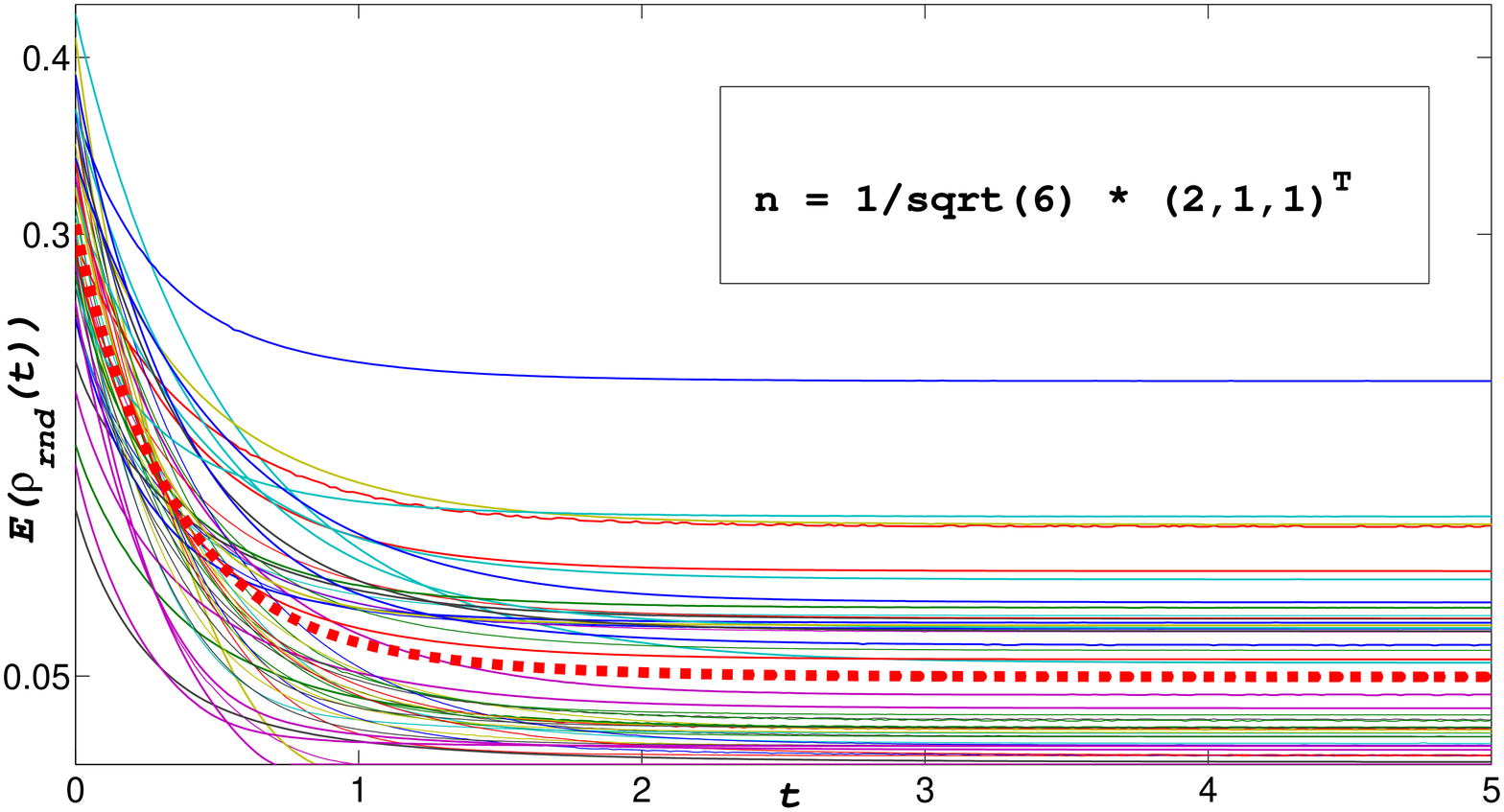}}
\centering
%\subfigure[]{\includegraphics[width=7.9cm]{FigRSEFLD1}}
%\subfigure[]{\includegraphics[width=7.9cm]{FigRSEFLD2}}
\caption{Entanglement freezing dynamics is also observed for random pure states mixed with white noise with $\alpha = 0.95$ and for two settingd of 
$\vec{n}$.}
\label{FIG:RSEF}
\end{figure}
Finally, in order to compare the dynamics, we generate some pure random states and mix them with while noise by taking $\alpha = 0.95$. 
Figure (\ref{FIG:RSEF}) depicts genuine negativity for such random states for two choices of $\vec{n}$. We can see that in each case majority of 
the random states exhibit entanglement freezing phenomenon. Hence it is a generic feature of this particular type of decoherence. We also plot the 
mean value of genuine negativity in each case denoted by red dashed line. This situation is similar to the special case of current dynamics 
with $\vec{n} = (0,0,1)^T$ where almost every asymptotic state is found to be genuinely entangled due to appearance of 
decoherence free spaces \cite{Ali-PRA81-2010}. However, it is important to note that there are no DFS for current settings of $\vec{n}$. 
It is the property of dynamics that causes this interesting feature of entanglement freezing dynamics for most of the states.

\begin{figure}[h]
\scalebox{2.20}{\includegraphics[width=5.5cm]{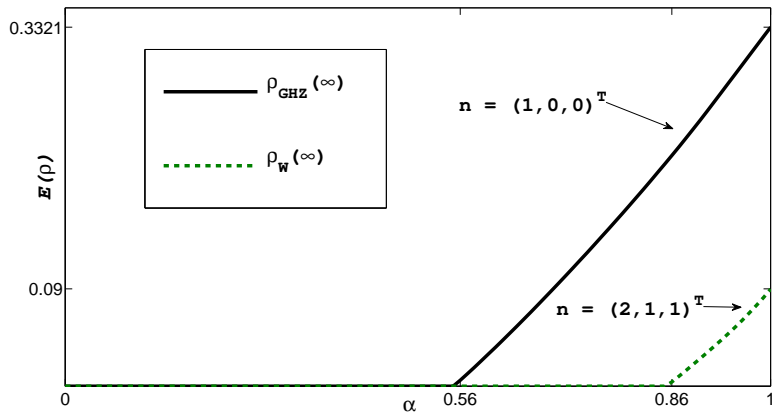}}
\caption{Genuine negativity is plotted against parameter $\alpha$ for asymptotic GHZ and W states. See text for details.}
\label{FIG:GHZWinf}
\end{figure}

A natural question arises that whether we can analyze the quantum states at infinity. With current dynamics, due to the fact that 
$\lim_{t\to \infty} \phi(t) = 0$, it is possible to check the asymptotic quantum states \cite{Carnio-PRL-2015}, and they are given as
\begin{eqnarray}
\rho(\infty) = \sum_i^N \, \Theta_i \, \rho(0) \, \Theta_i \, . 
\end{eqnarray}
However, due to presence of parameters $n_i$, it is quite cumbersome to write the asymptotic density matrix here. Instead, 
as we have seen from Figures~(\ref{FIG:WGHZn1} \& \ref{FIG:WGHZn2}) that either GHZ or W state exhibits freezing dynamics depending 
upon choice of $\vec{n}$, therefore we also analyze these states at infinity. 
% For an example, for an initial mixture of GHZ state with white noise, 
% the state at infinity for $\vec{n} = (1,0,0)^T$ is given as
% \begin{eqnarray}
% \rho_{GHZ}(\infty) = \left( 
% \begin{array}{cccccccc}
% \frac{2 + 3 \, \alpha}{16} & - \alpha/16 & - \alpha/16 & - \alpha/16 & - \alpha/16 & - \alpha/16 & - \alpha/16 & 5 \, \alpha/16 \\
% - \alpha/16 & \frac{2-\alpha}{16} & \alpha/16 & \alpha/16 & \alpha/16 & \alpha/16 & \alpha/16 & - \alpha/16 \\
% - \alpha/16 & \alpha/16 & \frac{2 - \alpha}{16} & \alpha/16 & \alpha/16 & \alpha/16 & \alpha/16 & - \alpha/16 \\ 
% - \alpha/16 & \alpha/16 & \alpha/16 & \alpha/16 & \frac{2-\alpha}{16} & \alpha/16 & \alpha/16 & -\alpha/16 \\
% - \alpha/16 & \alpha/16 & \alpha/16 & \alpha/16 & \frac{2-\alpha}{16} & \alpha/16 & \alpha/16 & - \alpha/16 \\
% - \alpha/16 & \alpha/16 & \alpha/16 & \alpha/16 & \alpha/16 & \frac{2-\alpha}{16} & \alpha/16 & - \alpha/16\\
% - \alpha/16 & \alpha/16 & \alpha/16 & \alpha/16 & \alpha/16 & \alpha/16 & \frac{2-\alpha}{16} & - \alpha/16\\
% 5\, \alpha/16 & - \alpha/16 & - \alpha/16 & - \alpha/16 & - \alpha/16 & - \alpha/16 & - \alpha/16 & \frac{2 + 3 \, \alpha}{16} 
% \end{array}
% \right)\,.
% \label{Eq:GHZn1inf}
% \end{eqnarray}
%Similarly, we have a density matrix for $\rho_W(\infty)$ for $\vec{n} = (2,1,1)^T/\sqrt{6}$. 
In Figure~(\ref{FIG:GHZWinf}), we plot genuine negativity for asymptotic GHZ and W states against parameter $\alpha$. It is known that 
$\rho_{GHZ}(0)$ is genuinely entangled iff $\alpha \geq 0.429$, whereas for $\rho_W(0)$ the limit is $\alpha \geq 0.479$. We can see that for 
all $\alpha > 0.56$, the asymptotic GHZ states are genuinely entangled, whereas for genuinely entangled asymptotic W states we must have $\alpha > 0.86$. 

\subsection{Four qubits}

\begin{figure}[h]
\scalebox{2.20}{\includegraphics[width=5.5cm]{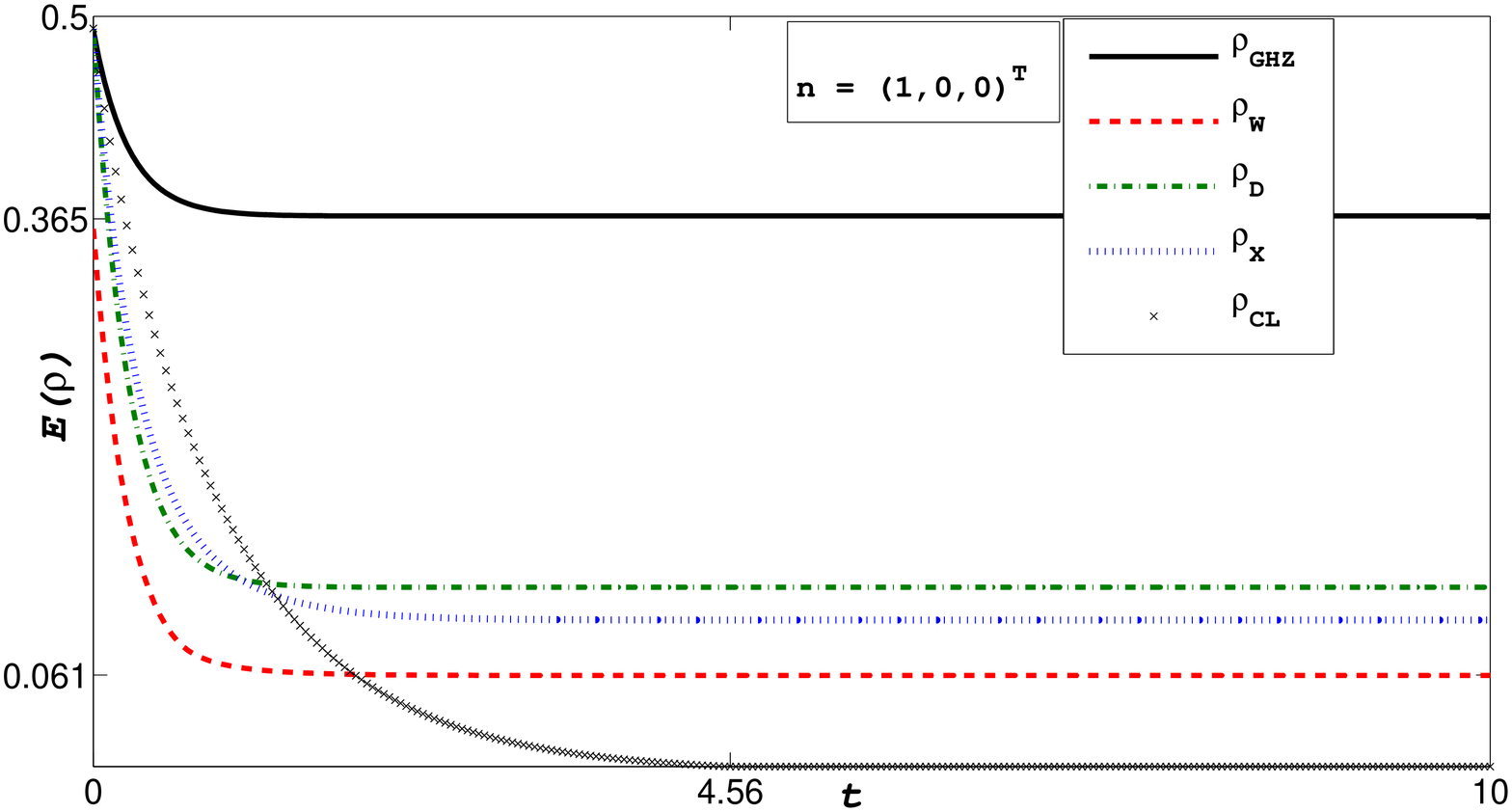}}
\caption{Genuine negativity is plotted against parameter $t$ for various four qubit entangled states. We take $\alpha = 0.99$. See text for details.}
\label{FIG:Q4s}
\end{figure}

In Figure (\ref{FIG:Q4s}), we plot the genuine negativity for various initial four qubits states already discussed in previous section. 
Interestingly, we see that for $\vec{n} = (1,0,0)^T$, $W_4$ state is quite robust as compared with $W_3$ which was quite fragile. $GHZ_4$ state is 
again robust just like $GHZ_3$. We can see that except for cluster state all other states exibit freezing entanglement dynamics. We have taken 
$\alpha = 0.99$.

\begin{figure}[h]
\scalebox{2.20}{\includegraphics[width=5.5cm]{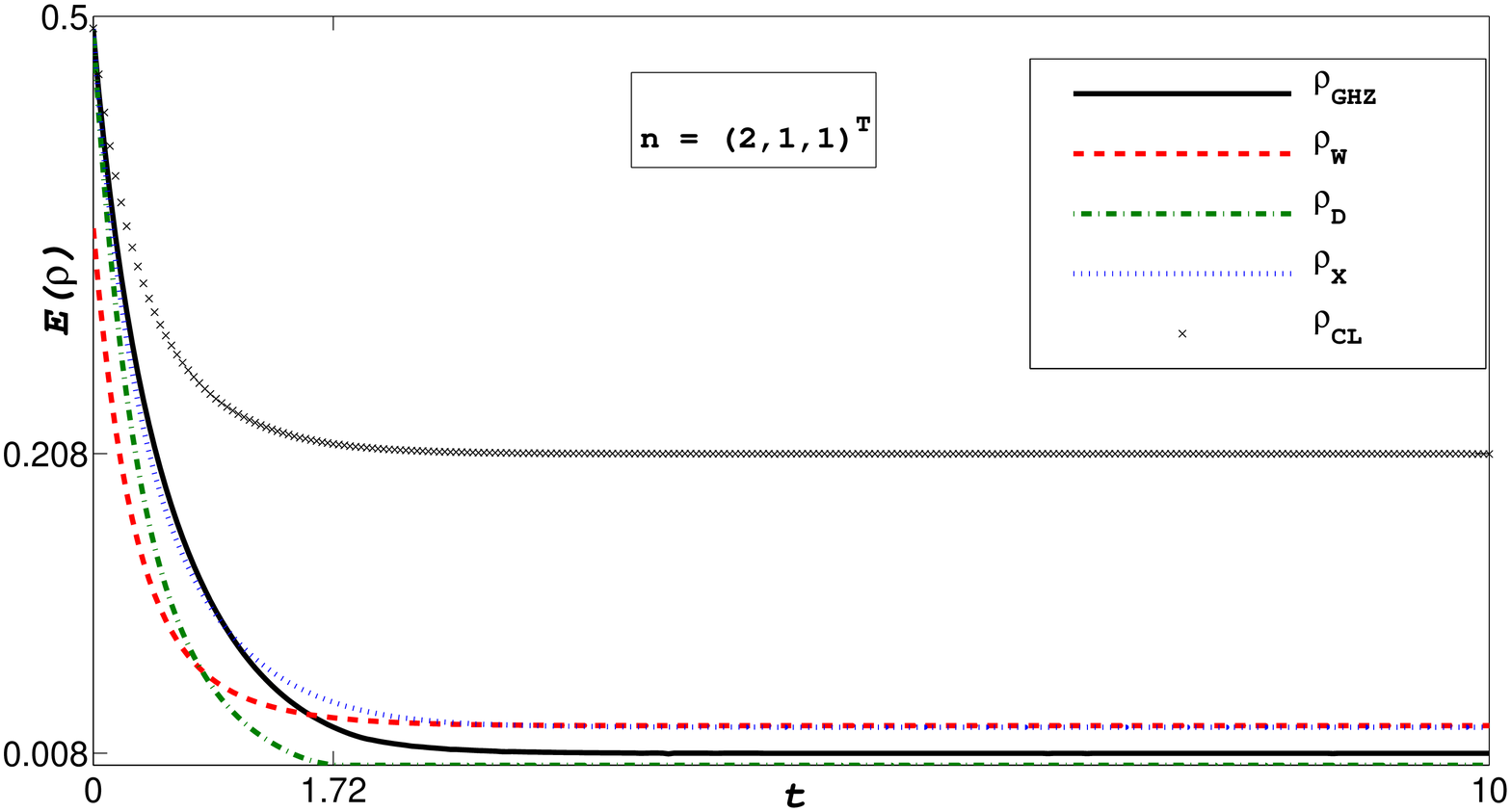}}
\caption{Genuine negativity is plotted against parameter $t$ for various four qubit entangled states. We take $\alpha = 0.99$. See text for details.}
\label{FIG:Q4n2s}
\end{figure}

Figure (\ref{FIG:Q4n2s}) depicts genuine negativity for four qubits states for another setting of $\vec{n} = 1/\sqrt{6} (2,1,1)^T$. In contrast with 
Figure (\ref{FIG:Q4s}), now the cluster state is quite robust. All states exhibit freezing dynamics except $\rho_{D}$, whose genuine negativity 
becomes zero at $\approx 1.72$.

We expect the similar behaviour of random pure states of four qubits mixed with white noise as for three qubits.
% \begin{figure}[h]
% \scalebox{2.20}{\includegraphics[width=5.5cm]{Fig4Qinf}}
% \caption{Genuine negativity is plotted against parameter $\alpha$ for asymptotic four qubit states. See text for details.}
% \label{FIG:Q4infn1}
% \end{figure}
% 
% \begin{figure}[h]
% \scalebox{2.20}{\includegraphics[width=5.5cm]{Fig4Qinfn2}}
% \caption{Genuine negativity is plotted against parameter $\alpha$ for asymptotic four qubit states. See text for details.}
% \label{FIG:Q4infn2}
% \end{figure}

\subsection{Finite time end of genuine nonlocality}

In this subsection, we show that genuine nonlocality undergoes finite time end even though the quantum states are genuinely entangled. For three qubits 
it is already known that Svetlichny inequality is maximally violated by $GHZ_3$ state and the maximum violation is equal to $4\sqrt{2}$, whereas for 
$W_3$ state the maximum violation can be ``$5 \sin\tilde{\theta} + \sin3\tilde{\theta} = 4.3546$'' for 
$\tilde{\theta} = 54.736^\circ$. The choice of measurement operators $M_i(M'_i)$ depends on initial state. For $GHZ_3$ state, we define 
$M_A \equiv \sigma_y$ and $M'_A \equiv \sigma_x$, the measurement operators for other two subsystems are defined with respect to the first by 
a rotation angle $\theta_K$ as
\begin{eqnarray}
\left(
\begin{array}{c}
M_K \\ 
M'_K\end{array}
\right)\, = R(\theta_K) \left( \begin{array}{c} M_A \\ M'_A \end{array} \right) \, , 
\label{Eq:MR}
\end{eqnarray}
where 
\begin{eqnarray}
R(\theta_K) = \left(
\begin{array}{cc}
\cos \theta_K & -\sin\theta_K \\ 
\sin\theta_K & \cos\theta_K 
\end{array}
\right)\, . 
\label{Eq:Rot}
\end{eqnarray}
Therefore the corresponding operators for three qubits are given as
\begin{eqnarray}
M_A &=& \sigma_y \otimes I_2 \otimes I_2 \, ,\nonumber \\
M'_A &=& \sigma_x \otimes I_2 \otimes I_2 \, ,\nonumber \\
M_B &=& I_2 \otimes [\cos\theta_B \, \sigma_y - \sin\theta_B \, \sigma_x ] \otimes I_2 \, ,\nonumber \\
M'_B &=& I_2 \otimes [\sin\theta_B \, \sigma_y + \cos\theta_B \, \sigma_x ] \otimes I_2 \, ,\nonumber \\
M_C &=& I_2 \otimes I_2 \otimes [\cos\theta_C \, \sigma_y - \sin\theta_C \, \sigma_x ] \, ,\nonumber \\
M'_C &=& I_2 \otimes I_2 \otimes [\sin\theta_C \, \sigma_y + \cos\theta_B \, \sigma_x ] \, .
\label{Eq:Mops}
\end{eqnarray}
It turns out that for $GHZ_3$ state mixed with white noise, the expression for $|\langle S\rangle|$ with $\vec{n} = (1,0,0)^T$ can be written as
\begin{eqnarray}
|\langle S\rangle|_{\rho_{GHZ}(t)} &=& \frac{\alpha \, (3 \, e^{-2 \, t} + 5)}{2} \, [\cos(\theta_B + \theta_C) 
- \sin(\theta_B + \theta_C)] \nonumber\\
|\langle S_{max}\rangle|_{\rho_{GHZ}(t)} &=& \frac{\alpha \, (3 \, e^{-2 \, t} + 5)}{\sqrt{2}} \, ,  
\end{eqnarray}
for $(\theta_B + \theta_C) = -\pi/4$. It is obvious that for $t = 0$, we have $|\langle S_{max}\rangle|_{\rho_{GHZ}(0)} = 4\sqrt{2} \, \alpha$, 
the maximum possible violation, whereas for $t \to \infty$, we have $|\langle S_{max}\rangle|_{\rho_{GHZ}(\infty)} = 3.54 \, \alpha$. This 
value is less than $4$ for the range of $\alpha$. On the other hand the corresponding expression for  $\vec{n} = 1/\sqrt{6} \, (2,1,1)^T$ also 
involves the sine and cosine terms with argument $(\theta_B + \theta_C) = -\pi/4$ for maximum value. The expression is given as
\begin{eqnarray}
|\langle S_2\rangle|_{\rho_{GHZ}(t)} &=& \frac{\alpha \, [20 + 1095 \, e^{- t} + 372 \, e^{-2 \, t} + 241 \, e^{- 3 \, t}]}{216 \, \sqrt{2}} \, , 
\end{eqnarray}
such that at $t= 0$, we have $|\langle S_2\rangle|_{\rho_{GHZ}(0)} = 4 \, \sqrt{2} \, \alpha$, the maximum possible violation and for $t \to \infty$, 
we have $|\langle S_2\rangle|_{\rho_{GHZ}(\infty)} = (5 \, \alpha)/(54 \, \sqrt{2})$, which far below than $4$ even for $\alpha = 1$. This means 
that these states lose their nonlocality at a finite time under current model of decoherence.

The measurement operators for $W_3$ state can be written as
\begin{eqnarray}
M_A &=& [\cos\tilde{\theta} \, \sigma_x + \sin\tilde{\theta} \, \sigma_z ] \otimes I_2 \otimes I_2 \, ,\nonumber \\
M'_A &=& [\cos\tilde{\theta} \, \sigma_x - \sin\tilde{\theta} \, \sigma_z ] \otimes I_2 \otimes I_2 \, ,\nonumber \\
M_B &=& I_2 \otimes [\cos\tilde{\theta} \, \sigma_x + \sin\tilde{\theta} \, \sigma_z ] \otimes I_2 \, ,\nonumber \\
M'_B &=& I_2 \otimes [\cos\tilde{\theta} \, \sigma_x - \sin\tilde{\theta} \, \sigma_z ] \otimes I_2 \, ,\nonumber \\
M_C &=& I_2 \otimes I_2 \otimes [\cos\tilde{\theta} \, \sigma_x + \sin\tilde{\theta} \, \sigma_z ] \, ,\nonumber \\
M'_C &=& I_2 \otimes I_2 \otimes [\cos\tilde{\theta} \, \sigma_x - \sin\tilde{\theta} \, \sigma_z ] \, 
\label{Eq:MopsW}
\end{eqnarray}
where $\tilde{\theta}$ defined earlier. For $W_3$ state mixed with white noise, the expression for $|\langle S\rangle|$ for $\vec{n} = (1,0,0)^T$ 
can be written as
\begin{eqnarray}
|\langle S\rangle|_{\rho_W(t)} &=& \frac{\alpha \, \sin\tilde{\theta} \, e^{-3 \, t}}{2} \, \big[\, 3 + 9 e^{2 \, t} 
+ \cos 2\tilde{\theta} \, (-3 + 7 \, e^{2 \, t})\,\big]\,, 
\end{eqnarray}
which is $|\langle S\rangle|_{\rho_W(0)} = 4.3546 \, \alpha$, the maximum violation at $t = 0$ and $|\langle S\rangle|_{\rho_W(\infty)} = 0$ 
for $t = \infty$. The corresponding expression for $\vec{n} = 1/\sqrt{6} \, (2,1,1)^T$ is given as
\begin{eqnarray}
|\langle S_2\rangle|_{\rho_W(t)} &=& 1.3 \, \alpha \, e^{-3 \, t} \, [0.444 + e^t][1.83 + e^t (-0.5134 + e^t)] \, , 
\end{eqnarray}
which is $|\langle S_2\rangle|_{\rho_W(0)} = 4.3546 \, \alpha$, the maximum violation at $t = 0$ and $|\langle S_2\rangle|_{\rho_W(\infty)} = 0$ 
for $t = \infty$. Hence the nonlocality of $W_3$ states turns out to be extremly fragile under collective dephasing.

As discussed in previous section, the genuine nonlocality quantified by Svetlichny inequality come to an end at a finite time for various 
settings of $\vec{n}$, although the states remain genuine entangled. This situation is similar to two qubit case where the nonlocality is lost 
at a finite time whereas the quantum states remain entangled \cite{Liu-arXiv}. 

% Figure (\ref{FIG:SGHZ}) shows the quantity $|\langle S \rangle|$ for GHZ states 
% with two settings of $\vec{n}$. As it was clear from its analytical expression, the inequality no longer violates the limit $4$ after a short 
% time and hence lose its nonlocality even for $\alpha \to 1$.
% \begin{figure}[h]
% \scalebox{2.20}{\includegraphics[width=5.7cm]{SNLGHZ}}
% \caption{Svetlichny inequality is plotted against $t$ for two settings of $\vec{n}$ for $GHZ_3$ states. We take $\alpha = 0.99$.}
% \label{FIG:SGHZ}
% \end{figure}
% 
% Figure (\ref{FIG:SW}) reflects the fragility of nonlocality of $\rho_W(t)$ states under collective dephasing. For both settings of $\vec{n}$, 
% the inequality becomes smaller than $4$ after a very short time. 
% \begin{figure}[h]
% \scalebox{2.20}{\includegraphics[width=5.7cm]{SNLW}}
% \caption{Svetlichny inequality is plotted against $t$ for two settings of $\vec{n}$ for $W_3$ states. We take $\alpha = 0.99$.}
% \label{FIG:SW}
% \end{figure}

The Svetlichny inequality for four qubits GHZ state mixed with white noise for $\vec{n} = (1,0,0)^T$ is given as
\begin{eqnarray}
|\langle S_{max} \rangle|_{\rho_{GHZ}(t)} &=& \frac{\alpha \, (19 + 12 \, e^{-2 \, t} + e^{-4 \,t})}{2 \,\sqrt{2}} \, ,  
\end{eqnarray}
where we have taken $\theta_B + \theta_C + \theta_D = -\pi/4$. At $t = 0$, we have $|\langle S_{max} \rangle|_{\rho_{GHZ}(0)} = 8 \, \sqrt{2} \, \alpha$, 
the maximum violation, whereas for $t \to \infty$, we have $|\langle S_{max} \rangle|_{\rho_{GHZ}(\infty)} = (19 \alpha)/(2\sqrt{2}) < 8$. So we again 
see that even though the state is genuinely entangled but its genuine nonlocality is lost at finite time. Similarly, the expression for 
$\vec{n} = (2,1,1)^T/\sqrt{6}$, we have 
\begin{eqnarray}
|\langle S_2 \rangle|_{max} &=& \frac{\alpha \, (329 + 6440 e^{-t} + 2996 e^{-2 \, t} + 3352 \, e^{-3 \, t} 
+ 707 e^{-4 \,t})}{864 \,\sqrt{2}} \, .  
\end{eqnarray}
This value is maximum at $t = 0$ and becomes smaller than $8$ at a finite time, means end of genuine nonlocality.

%%%%%%%%%%%%%%%%%%%%%%%%%%%%%%%% 
\section{Discussions and Summary} \label{Sec:conc}
%%%%%%%%%%%%%%%%%%%%%%%%%%%%%%%% 

We analyzed the dynamics of genuine multipartite entanglement and genuine nonlocality for three and four qubits under general model of 
collective dephasing. We found that for certain directions of $\vec{n}$, entanglement of $GHZ_3$ and $W_3$ states first decay upto a certain 
value and exhibit freezing dynamics afterwards. This is an interesting feature as quantum states are changing but their entanglement is locked to a 
specific value. We pointed out that entanglement {\it freezing} is different feature than {\it time-invariant} entanglement. 
We then studied the dynamics for random pure states mixed with white noise and found that genuine entanglement in most of these states also decay 
initially to some value and later exhibit stationary entanglement. We also observed this {\it freezing} dynamics of entanglement for various four 
qubits genuine entangled states. On the other hand the genuine nonlocality of these quantum states measured by Svetlichny inequality suffer a finite 
time end even though the states remain genuine entangled. These observations are similar to such studies for two qubits case. One of the future avenue 
would be to look for the time-invariant feature for quantum nonlocality.

\section*{Acknowledgments}
The author is grateful to Edoardo G. Carnio for helpful discussions and Otfried G\"uhne for his useful correspondence. The author is also thankful to 
A. R. P. Rau for reading the manuscript and to referee for his/her positive comments.

\end{document}